\documentclass[onecolumn]{webofc}

\usepackage[varg]{txfonts}

\usepackage{hyperref}

\usepackage{booktabs}

\usepackage{titlesec}

\titleformat{\paragraph}[runin] 
  {\normalfont\normalsize\bfseries}{}{0em}{}
\titlespacing{\paragraph}
  {0pt}{0.5em}{0.5em} 

\usepackage[belowskip=-5pt,aboveskip=5pt]{caption}
\begin{document}

\title{Performance Portable Monte Carlo Particle Transport on\\Intel, NVIDIA, and AMD GPUs}

\author{
\firstname{John} \lastname{Tramm}\inst{1}\fnsep\thanks{\email{jtramm@anl.gov}}
\and
\firstname{Paul} \lastname{Romano}\inst{1}
\and
\firstname{Patrick} \lastname{Shriwise}\inst{1}
\and
\firstname{Amanda} \lastname{Lund}\inst{1}
\and
\firstname{Johannes} \lastname{Doerfert}\inst{2}
\and
\firstname{Patrick} \lastname{Steinbrecher}\inst{3}
\and
\firstname{Andrew} \lastname{Siegel}\inst{1}
\and
\firstname{Gavin} \lastname{Ridley}\inst{4}
}

\institute{Argonne National Laboratory, USA
\and
           Lawrence Livermore National Laboratory, USA
\and
            Intel Corporation, USA
\and
           Massachusetts Institute of Technology, USA
          }

\abstract{
 OpenMC is an open source Monte Carlo neutral particle transport application that has recently been ported to GPU using the OpenMP target offloading model. We examine the performance of OpenMC at scale on the Frontier, Polaris, and Aurora supercomputers, demonstrating that performance portability has been achieved by OpenMC across all three major GPU vendors (AMD, NVIDIA, and Intel). OpenMC's GPU performance is compared to both the traditional CPU-based version of OpenMC as well as several other state-of-the-art CPU-based Monte Carlo particle transport applications. We also provide historical context by analyzing OpenMC's performance on several legacy GPU and CPU architectures. This work includes some of the first published results for a scientific simulation application at scale on a supercomputer featuring Intel's Max series ``Ponte Vecchio'' GPUs. It is also one of the first demonstrations of a large scientific production application using the OpenMP target offloading model to achieve high performance on all three major GPU platforms. 
}

\maketitle

\section{Introduction}

The recent decade has seen a significant increase in interest in using the Monte Carlo (MC) method for routine reactor design and engineering simulation work. The extremely high computational cost of using the MC method for full core simulations makes these problems ideal use cases for exascale class supercomputers. Therefore, there is a great need for MC applications that can run and scale efficiently---and in a portable manner---on all types of modern GPU-based systems.

\subsection{Performance Portability}

The world's first exascale supercomputer, Oak Ridge National Laboratory's ``Frontier'', features MI250X GPUs from AMD. The second exascale computer, Argonne National Laboratory's ``Aurora'', features Intel ``Ponte Vecchio'' (PVC) Max 1550 GPUs. These first entries into the exascale era of supercomputing are not notable due to their usage of GPUs (as most leadership class supercomputers in recent years have used GPUs)---they are, however, notable in that they do not use NVIDIA GPUs. This sudden diversification in the GPU market has immediate implications for scientific simulation application development in that NVIDIA's proprietary CUDA programming model is not supported on many leadership class systems leading to a crisis of portability for established GPU codes. There are a variety of portable GPU programming models (e.g., OpenMP target offloading, HIP, SYCL/DPC++, Kokkos, RAJA, OpenCL, OCCA, and Legion, just to name a few), yet all feature differing levels of support across vendors and are at different stages of maturity. Most importantly, as many of the models are very new compared to CUDA, it is still an open question as to whether or not portable models are capable of providing full performance portability for HPC production simulation applications at scale.

Given the near-universal requirement for simulation applications to run in a portable manner on GPU, there has been a push in recent years for legacy CPU-based applications to be ported to GPU and re-optimized to run efficiently on GPU architectures~\cite{Tramm2022,hamilton19annals,choi,Mckinley2019}. MC neutral particle transport is one such field of simulation science; many MC codes have seen decades of development and optimization for CPU-based systems, and their stochastic (direct simulation) approach makes them inherently difficult to optimize on most computer architectures. As such, great effort in recent years has been focused on porting and optimizing MC particle transport applications for GPU with the goal of bringing previously impractical simulation problems within reach on GPU-based exascale systems.

\subsection{Application}
\label{sec:openmc}

In this paper, we study a specific neutral particle transport application, OpenMC~\cite{romano-2015-ane1}. OpenMC is a continuous-energy MC particle transport application that has demonstrated excellent performance and scalability on CPU-based systems. In addition to being an open source application, OpenMC also provides a host of advanced modeling and simulation capabilities including depletion, advanced geometry representations, on-the-fly Doppler broadening, and multigroup cross section generation. Recently, it has also been ported to GPU using the OpenMP target offloading programming model, although so far only preliminary results have been published for a subset of single-GPU architectures (NVIDIA and AMD only), and no multi-GPU or multi-node scaling studies have yet been reported~\cite{Tramm2022}. More information about the algorithms and optimizations utilized in OpenMC's OpenMP offloading port is given in~\cite{Tramm2022}. 

In the present work, we analyze OpenMC's performance on a wide variety of GPU architectures, including GPUs manufactured by NVIDIA, AMD, and Intel. Additionally, we utilize OpenMC's MPI domain replication capabilities~\cite{Romano2012} to examine its scalability on larger GPU-based supercomputers. While OpenMC's domain replicated communication patterns have historically been relatively inexpensive on CPU-based supercomputers, GPU-based nodes may be capable of much faster per-node particle tracking rates that may result in a large relative increase in the communication cost. As such, it is important to examine OpenMC's scalability to evaluate communication costs on GPU-based systems.

Beyond our work on the OpenMP offloading port of OpenMC, there is a growing body of research porting MC particle transport applications to GPUs. For instance,  Shift~\cite{hamilton19annals}, PRAGMA~\cite{choi}, and Mercury~\cite{Mckinley2019} all utilize CUDA to leverage NVIDIA GPUs. Shift has also been ported to the HIP programming model in order to utilize AMD GPUs. There have also been several recent efforts to port smaller portions of OpenMC into the CUDA language~\cite{ridley2021,openmc_optix}. The present work regarding OpenMC differs from all these efforts in a number of ways, the most notable of which is the usage of the OpenMP target offload programming model allowing it to utilize GPUs from all three major GPU vendors as well as CPUs within a single codebase. While OpenMC's theoretical portability advantage is clear, in this work, we will investigate if the OpenMP offloading compiler toolchains are mature enough to deliver true performance portability. We also note that the OpenMP offloading port of OpenMC currently exists as a separate, open source fork~\cite{openmc_exasmr} as it has not yet been merged together with the main development branch of OpenMC~\cite{openmc_main}, although this may be pursued in the near future in light of the compelling speedups observed in this analysis.

\section{GPU-Specific Optimizations and Configurations}

Most of the key algorithmic changes and optimizations made to OpenMC to support efficient GPU execution are described in~\cite{Tramm2022}. However, this previous work was limited to NVIDIA and AMD GPUs, so we extend our discussion in this section to reflect new experiences in optimizing OpenMC for Intel GPUs as well as other optimizations we have developed since the publication of~\cite{Tramm2022}.

\paragraph{Event-Based Parallelism} Following previous work in the field~\cite{hamilton19annals}, the OpenMP offloading port of OpenMC uses ``event-based'' parallelism when executing on GPU and traditional ``history-based'' parallelism when executing on CPU. In the history-based method for MC, each particle is handled ``birth to death'' in a sequential fashion by a single thread, with parallelism expressed over fully independent histories.  In event-based mode in OpenMC, queues are maintained to decide which event to launch next, with each kernel call executing a narrowly defined operation (e.g., collision, cross section lookup, ray trace) over all particles at once. Analyses performed previously in~\cite{Tramm2022} showed that event-based parallelism is typically around 6$\times$ faster than history-based parallelism on GPU, while the opposite is true on CPU, where history-based parallelism is typically around 4$\times$ faster than event-based mode. These findings remain true for Intel GPUs.

\paragraph{Sorting} Periodically sorting particles has been found to greatly improve performance on all three GPU vendor architectures. Particles are sorted by the type of material they are traveling through and by energy before the cross section lookup kernel is executed. As described in~\cite{Tramm2022}, this sorting operation results in extreme memory bandwidth efficiency gains for the cross section lookup kernel (typically the most expensive part of the simulation), and we confirm that this sorting operation is also advantageous on Intel GPUs. Sorting a large buffer efficiently on device, however, is not an easy operation to code efficiently by hand (regardless of programming model). To this end, we link to vendor provided (GPU-specific) sorting libraries (CUDA Thrust, ROC Thrust, and Intel oneDPL), which is essentially the only area of vendor specialization in the OpenMP offloading port of OpenMC.

\paragraph{Tallying} To accelerate tallying of macroscopic quantities, we have added the ability for OpenMC to detect when certain depletion related tallies are in use and to assemble macroscopic cross section data as part of its normal cross section lookup kernel. While this marginally increases the cost of all cross section lookups during the active batches of OpenMC, the cost is outweighed by the savings in the tallying kernel as it removed the need to loop over all nuclides within a material to reassemble the macroscopic cross section when tallying. For the largest run in this paper (2048 nodes of the Frontier supercomputer using MI250X GPUs), an inactive batch rate of 1.65 billion particles/sec was observed, with the active batch rate being only about 29\% lower at 1.27 billion particles/sec.

\paragraph{In-Flight Particles} In the event-based mode of OpenMC, the maximum number of particles that are allowed to be ``in-flight'' at any one point in time can be controlled by the user. This quantity is separate from the number of particles per batch, so it does not affect any numerical aspect of the simulation but does ensure that the memory requirements for a kernel call stay below the available device memory. Given the need for massive parallelism on GPU architectures, all three GPUs see performance gains as this value is increased up to about one million particles. We found that NVIDIA GPUs tended to also see incremental performance gains as more particles in-flight were added past one million. Conversely, all AMD and Intel GPUs we tested saw performance degradation past the peak of around one million. Thus, we pinned the number of particles in-flight to 1.125 million for all of our runs on Intel GPUs, and to 1 million on AMD GPUs, but extended this value up to the maximum allowed by memory on NVIDIA GPUs. Further performance gains are also observed on all three GPU architectures as more overall particles per batch are added, resulting in dynamic refilling of in-flight particles and amortizing the cost of long-tailed particle histories.

\paragraph{GPU Streams} A surprising result in adapting OpenMC for Intel GPUs was the discovery that OpenMC tends to perform better with multiple GPU streams operating on GPU devices concurrently. Launching of concurrent GPU streams was easily accomplished by simply assigning multiple MPI ranks to each GPU, rather than adding another layer of parallelism via OpenMP. On NVIDIA GPUs, this required launching the NVIDIA multi-process service to allow for interleaving of streams via MPI on GPU. On Intel GPUs, this required launching the application in 4 compute command streamer (CCS) mode. We found that AMD was optimal with two MPI ranks per non-uniform memory access (NUMA) GPU domain, NVIDIA was optimal with four or eight MPI ranks per GPU depending on memory availability, and that Intel was optimal with four MPI ranks per NUMA GPU domain. While this was a modest optimization on AMD and NVIDIA architectures (resulting in a 24\% speedup on the MI250X and 27\% speedup on the A100 on a depleted fuel test problem), the optimization was highly impactful on Intel GPUs, where a 75\% speedup was observed compared to running only a single MPI rank per tile.

\section{CPU Baseline and Single GPU Performance Results}
\label{sec:code_compare}

As we will not make use of any analytical performance model in this study to assess the performance of the GPU version of OpenMC, we are limited in comparing its performance against execution on CPU. While there are other GPU-based MC codes (e.g., Shift~\cite{hamilton19annals}, PRAGMA~\cite{choi}, and Mercury~\cite{Mckinley2019}), access to these applications is tightly controlled. However, we did have access to the Serpent 2.1.31~\cite{serpent}, MCNP 6.2~\cite{mcnp_62}, and SCONE~\cite{Kowalski2021} CPU codes; providing CPU baselines with these applications helps fully contextualize the performance of OpenMC on GPU. For simplicity, we chose a standard depleted pincell problem with 251 fuel nuclides. Each code was compiled with its optimization flags enabled with the GNU 9.4.0 compiler and run on a dual-socket Xeon Platinum 8260 node with 48 cores total. In some cases, certain algorithmic optimization options were not enabled despite their availability, as they are not applicable for large scale reactor simulation problems. Specifically, the full grid unionization~\cite{Leppanen2009} feature was not enabled in Serpent and SCONE. While this feature would have resulted in significant speedup for the pincell problem, it requires $\mathcal{O}(10)$ GB of memory per material, which is not feasible for full core depletion simulation problems that may have hundreds of thousands of unique material regions. In the absence of full grid unionization, alternative lookup acceleration methods (e.g., double indexing) were enabled in each code where available. Delta-tracking was explicitly disabled to enforce ray tracing (also known as surface tracking) for all applications to provide an ``apples-to-apples'' tracking rate FOM performance comparison as OpenMC currently only supports surface tracking. Additionally, as the tallying strategies used by the different codes tend to vary much more widely than the general particle tracking routines, only inactive batch performance (where no tallying is done) is compared so as to maintain the simplest basis of comparison.

\begin{figure}[htbp]
  \centering
  \includegraphics[width=0.6\linewidth]{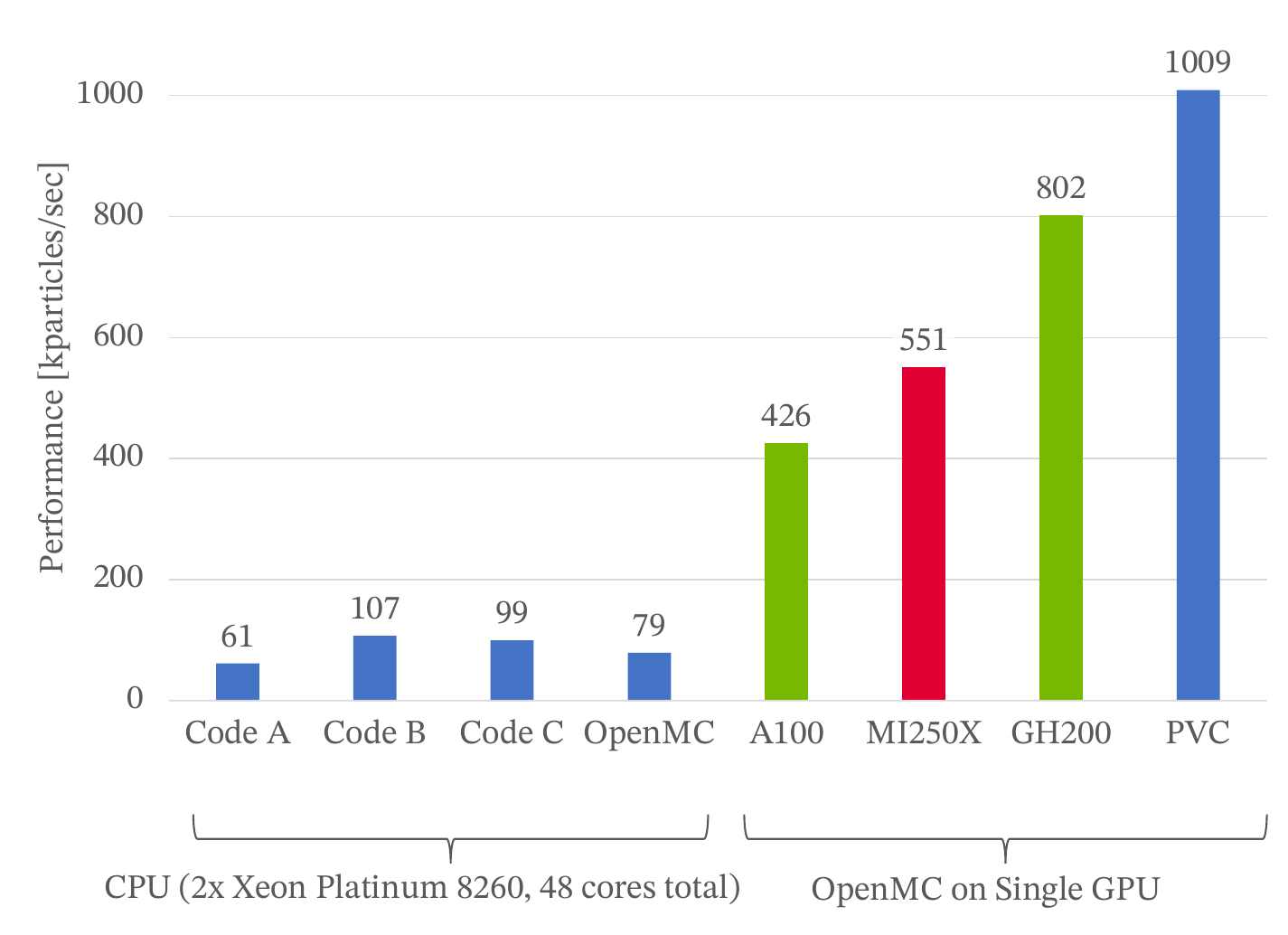}
  \caption{Comparison between other state-of-the-art Monte Carlo particle transport codes and OpenMC on a depleted pincell benchmark problem. Performance is measured for inactive batches.}
    \label{fig:code_compare}
\end{figure}

 We present the CPU performance results from Serpent, MCNP, and SCONE in \autoref{fig:code_compare} in an anonymized fashion to avoid being taken out of context given the limited nature of this comparison. These codes are listed as A, B, and C (in no particular order). While the OpenMP offloading port of OpenMC utilizes a single codebase for both CPU and all GPUs, the code was run using algorithms that maximized the performance of each architecture (i.e., history-based on CPU, and event-based on GPU). We also collected results for OpenMC on the pincell problem for a number of GPU systems for comparison (listed in \autoref{tab:supercomputers}), as well as an NVIDIA Grace Hopper GH200 GPU at Argonne National Laboratory. The results shown in \autoref{fig:code_compare} indicate that the CPU version of OpenMC compares very well with other state-of-the-art MC transport applications for this type of simulation problem. All four CPU codes have relatively similar performance levels (being 9--17$\times$ slower than the fastest GPU tested), and OpenMC on CPU is within about 35\% in performance of the fastest CPU code that we tested. Thus, it is reasonable to conclude that OpenMC's CPU performance is in line with the state-of-the-art, and that other MC applications might expect similar levels of speedup on GPU as observed with OpenMC.

\section{Supercomputer Performance Results}

\subsection{Simulation Test Problem}
\label{sec:problems}

OpenMC was part of the small modular reactor (SMR) simulation subproject ``ExaSMR'' in the broader Exascale Computing Project (ECP), which was a US DOE project. The goal of the ExaSMR project was to enable extreme-fidelity simulation of SMRs, which have been a recent focus of engineering development. OpenMC's role in ExaSMR was to serve as a MC particle transport solver in a multiphysics environment. Part of the ExaSMR project was to define a representative challenge problem. The selected challenge problem is a realistic SMR geometry that represents the NuScale reactor design. The SMR problem is defined with the goal of depletion calculations in mind, where hundreds of nuclides are present in the reactor fuel, and isotopic concentrations are maintained for hundreds of thousands of unique fuel burnup regions throughout the reactor core. While the SMR problem is defined in-depth in~\cite{ecp-se-08-43}, we note that in aggregate this problem featured 195,360 unique fuel material definitions and tally regions, with 244 nuclides in each fuel material, and 281 nuclides in the simulation overall. This type of simulation is therefore highly challenging given the expense of looking up fuel cross sections for hundreds of nuclides as particles transport through the system.

The relevant figure of merit (FOM) for the performance of the SMR simulation is the particle tracking rate (given in particles/sec) for the active phase of the simulation when tallies are being performed and statistics on final unknowns are being accumulated. Advanced features in OpenMC, such as probability tables and $S(\alpha,\beta)$ thermal scattering, are enabled to maximize fidelity and computational challenge. Additionally, the multipole cross section lookup method~\cite{forget2014} (with an advanced method for representing the Faddeeva function~\cite{forget2022}) in OpenMC is utilized to provide on-the-fly Doppler broadening, along with the relative velocity sampling (RVS) method~\cite{romano2018} for free gas elastic scattering.

\subsection{Test Systems}
\label{sec:systems}

To investigate the performance portability of OpenMC at scale we utilize a variety of modern leadership class supercomputing systems as detailed in \autoref{tab:supercomputers}. The Frontier supercomputer at Oak Ridge National Laboratory (ORNL) features a node design using MI250X GPUs developed by AMD. The Aurora supercomputer at Argonne National Laboratory (ANL) features a node design using Ponte Vecchio (PVC) Max 1550 GPUs developed by Intel. The third system, Polaris, is a pre-exascale NVIDIA A100 GPU-based supercomputer at ANL. While the A100 features a more traditional monolithic GPU design, both the MI250X and the PVC feature ``chiplet'' designs wherein each package is split into two different non-uniform memory access (NUMA) subdomains (which are called graphics compute dies (GCDs) by AMD and tiles by Intel). In this paper, we will consistently label the MI250X and the PVC as each being one GPU, where a single GPU is composed of either two MI250X GCDs or two PVC tiles.

\begin{table*}[]
\centering
\caption{Overview of HPC Systems}
\label{tab:supercomputers}
\begin{tabular}{@{}llllll@{}}
\toprule
System &
  \begin{tabular}[c]{@{}l@{}}Node\\ Count\end{tabular} &
  \begin{tabular}[c]{@{}l@{}}GPU\\ Architecture\end{tabular} &
  \begin{tabular}[c]{@{}l@{}}Number of\\ GPUs\\ per node\end{tabular} &
  \begin{tabular}[c]{@{}l@{}}Host\\ Architecture\end{tabular} &
  \begin{tabular}[c]{@{}l@{}}OpenMP\\ Compiler\end{tabular} \\ \midrule
Polaris                        & 560 & NVIDIA A100               & 4 & 1x EPYC 7543P & LLVM  \\
Frontier & 9,402 & AMD MI250X                & 4 & 1x EPYC 7A53  & LLVM  \\
Aurora  & 10,624 & Intel PVC Max 1550 & 6 & 2x Xeon 9470C & Intel oneAPI \\ \bottomrule
\end{tabular}
\end{table*}

\subsection{Single Node Performance}

\begin{figure*}[htbp]
  \centering
  \includegraphics[width=0.6\linewidth]{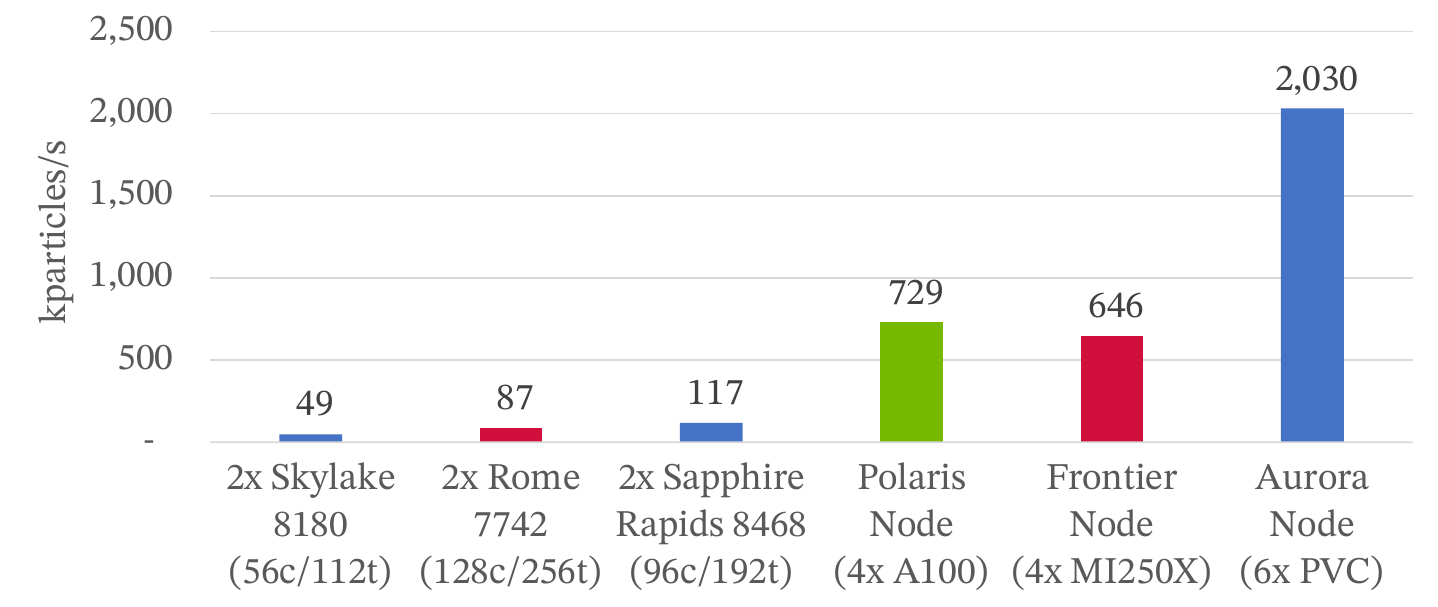}
  \caption{Performance comparison of OpenMC on various node architectures. Performance is given in terms of active batch (i.e., with tally) particle rates on a depleted small modular reactor problem with 195k material regions.}
    \label{fig:node_compare}
\end{figure*}

We begin our performance analysis by evaluating the node-level performance of OpenMC on the Aurora, Frontier, and Polaris systems as well as a variety of dual-socket CPU nodes to serve as a baseline. As shown in \autoref{fig:node_compare}, we can see that OpenMC runs efficiently on A100, MI250X, and PVC GPUs, allowing for significant gains over a variety of different dual-socket CPU nodes.

\subsection{Many Node Performance}

As discussed in more depth in \autoref{sec:openmc}, the MPI communication patterns used by OpenMC typically result in very low communication costs on CPU-based architectures. However, given the massive improvements in node-level particle tracking rates that GPUs offer (as evidenced by \autoref{fig:node_compare}), it is highly important to evaluate OpenMC at scale to determine if communication costs have become relatively more expensive. To this end, we perform weak scaling studies on the Frontier, Aurora, and Polaris systems. While the Polaris and Frontier studies use a large fraction of the available nodes on each system, we were only able to run on 325 nodes of the Aurora system given that it is still in pre-production mode and system availability was tightly limited due to acceptance testing taking priority.

\begin{figure}[htbp]
  \centering
  \includegraphics[width=0.6\linewidth]{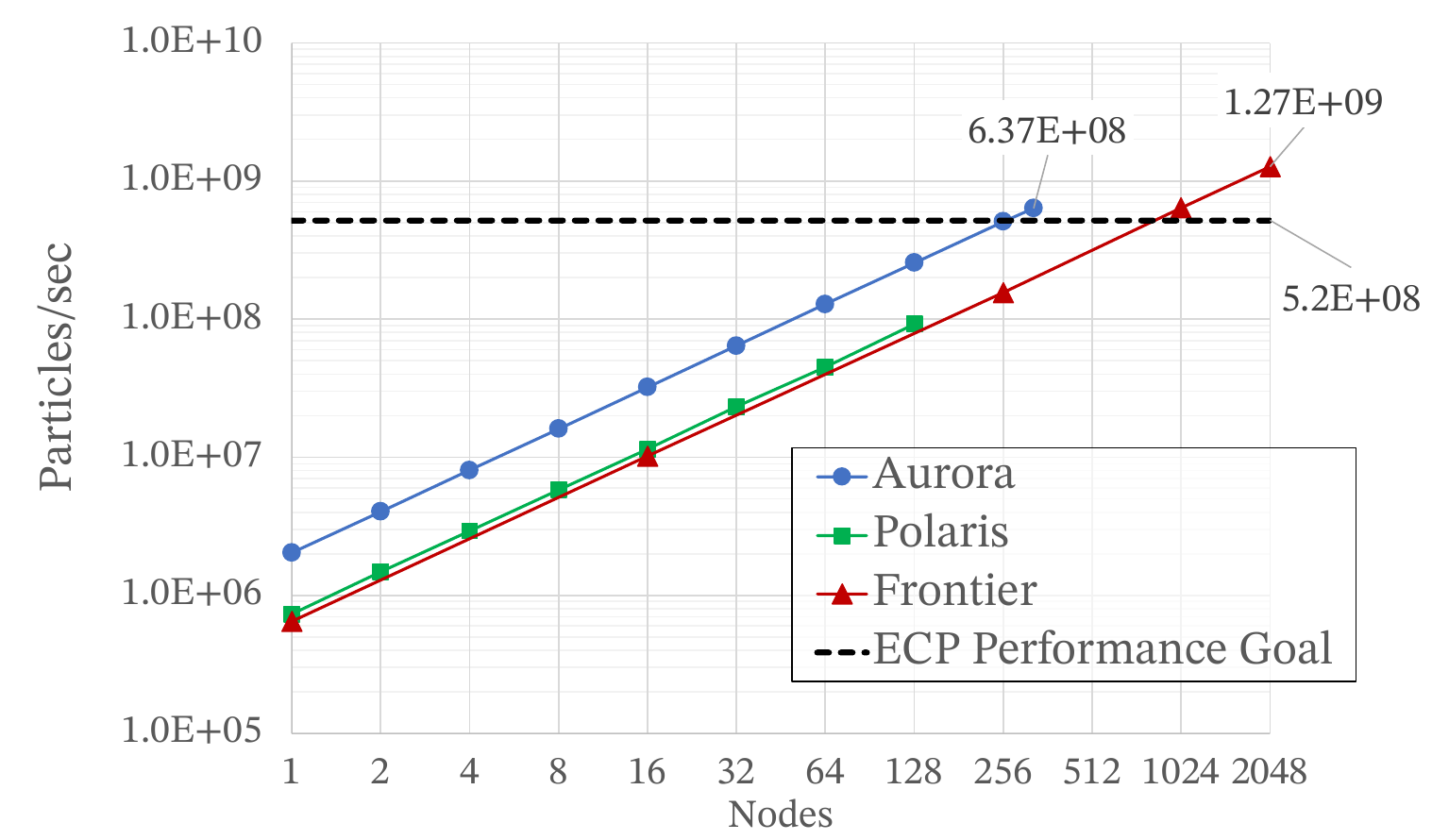}
  \caption{Weak scaling (where the problem size per GPU remains constant) performance of OpenMC on various architectures. Performance is given in terms of active batch (i.e., with tally) performance on a depleted small modular reactor problem with 195k material regions.}
    \label{fig:weak_scaling_absolute}
\end{figure}

Our weak scaling results, shown in \autoref{fig:weak_scaling_absolute}, indicate that the total particle tracking rate appears to increase nearly linearly on all three systems. For the largest runs, performance on Aurora, Polaris, and Frontier achieved 97\%, 96\%, and 99\% weak scaling efficiencies, respectively.

To give some sense of the magnitude of the performance numbers delivered by OpenMC at scale on these systems, we need only look at the performance goal used by the ExaSMR ECP project. The performance goal was to achieve a 50$\times$ speedup over state-of-the-art full system execution when ECP began. The baseline was set by using the Shift~\cite{hamilton19annals} MC code as run on the ORNL Titan supercomputer, which utilized NVIDIA K20X GPUs. The performance figure of merit (FOM) achieved by Shift's GPU baseline run on Titan (projected if using all of its 18,688 nodes) for a depleted SMR simulation was 10.39 million particles/sec~\cite{ECP-SE-08-119}. As shown in \autoref{fig:weak_scaling_absolute}, OpenMC was able to exceed the goal of a 50$\times$ speedup over Titan using only 325 nodes of Aurora or 1024 nodes of Frontier. We also note that OpenMC is the first MC application that we are aware of to achieve 1 billion particles per second on this class of problem (featuring depleted fuel with hundreds of thousands of unique material regions).

As discussed in~\cite{Tramm2022}, the OpenMP offloading port was engineered with reproducibility in mind such that results for serial CPU execution before the porting effort began and results at scale on GPU-based supercomputers are identical (although floating-point non-associativity effects are inevitable at scale). However, in the present work, we did implement a strategy for using multi-temperature data sets that resulted in divergence between history-based and event-based modes due to differences in stochastic temperature interpolation strategies. GPU vs. CPU solutions are nonetheless statistically identical. In future work, we aim to extend full reproducibility to multiple temperature data sets.

\section{Historical Performance}

The analyses so far in this paper have focused on the performance of modern HPC CPU and GPU resources. While these analyses have shown that OpenMC's GPU performance is many times faster than its CPU performance on current systems, they do not shed light on longer term trends in computer architecture. That is, is the performance gap between OpenMC on GPU and CPU relatively static over time, or is it becoming wider or smaller with each new architectural generation that is released? To answer this important question, we ran OpenMC on a number of legacy CPU and GPU architectures that were available on the Joint Laboratory for System Evaluation (JLSE) cluster at ANL. Legacy GPUs date back to 2012's K20X (notably, the GPU architecture that was utilized by the Titan supercomputer at ORNL when ECP was just beginning). We also selected legacy CPUs that were premier class, high-end dual-socket CPU nodes of their era. The release dates for all architectures are taken from~\cite{techpowerup}. We utilize the same depleted pincell problem that was used in \autoref{sec:code_compare}.

\begin{figure*}[htbp]
  \centering
  \includegraphics[width=0.6\linewidth]{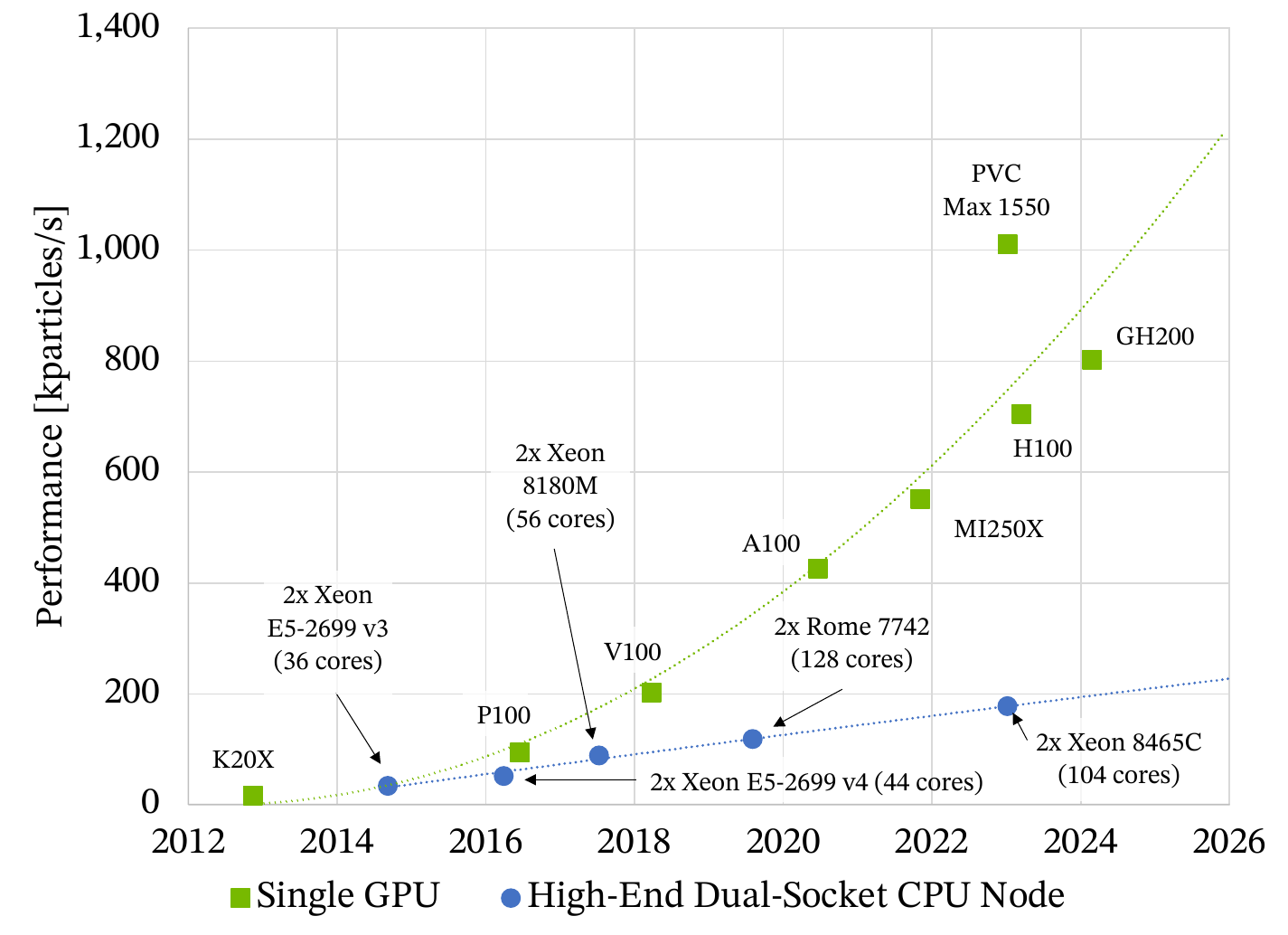}
  \caption{Performance of OpenMC on legacy CPU and GPU architectures by release date for a depleted pincell benchmark problem. }
    \label{fig:historical}
\end{figure*}

The results of our historical analysis, shown in \autoref{fig:historical}, reveal a long term trend of an exponentially widening performance gap between single GPU and dual-socket CPU performance. For instance, it took about 5 years for CPU core counts to double between the dual-socket Xeon 8180M (Skylake) node and the dual-socket Xeon 8465C (Sapphire Rapids) node, with only about 10\% gains in performance per core during that timeframe. Thus, the doubling time for OpenMC's performance on CPU architectures that we sampled was about 5 years. Conversely, in the 4 years between the release of the NVIDIA P100 and the NVIDIA A100, OpenMC's performance improved by roughly a factor of 4$\times$---a doubling time of only 2 years.

\section{Conclusions}

The purpose of this paper was to determine how well the MC neutral particle transport application, OpenMC, scaled on a variety of GPU-based systems. In particular, some of our key goals were to understand how well the application fared on all types of GPUs, including new Ponte Vecchio GPUs made by Intel, and whether or not it was possible for a large, complex simulation application like OpenMC to achieve robust performance portability via the OpenMP target offloading model.

Our results were surprising in a number of ways. First of all, despite this being Intel's first entry into the HPC GPU market, the Intel PVC Max 1550 greatly outperformed both the NVIDIA and AMD offerings. The Intel PVC was found to be about 2.3$\times$ faster than the A100, 1.2$\times$ faster than the GH200, and over 1.8$\times$ faster than the AMD MI250X on the depleted fuel problems we tested.

Another key finding was that OpenMC was in fact capable of running efficiently on GPUs from Intel, NVIDIA, and AMD using the OpenMP target offloading model. GPU node resources on the Aurora, Frontier, and Polaris systems yielded massive speedups as compared to running OpenMC on CPU-only node designs. For instance, compared with a state-of-the-art 96 core Sapphire Rapids dual-socket CPU node, a single Aurora GPU node was over 17$\times$ faster.

Another notable finding was that all systems consistently delivered >96\% weak scaling efficiency at scale on hundreds or thousands of nodes. These results demonstrate that all three machines had networks that were operating efficiently and were allowing for near-perfect intranode scaling (to multiple GPUs within the same node) and internode scaling (to many nodes) via MPI domain replication.

Overall, the results within this paper are notable as they are some of the first published results demonstrating the performance capabilities of Intel's PVC GPUs and the capability of the OpenMP target offloading model to deliver performance for large scientific applications at scale on GPUs from all major vendors. These results are also notable given the massive boost in MC particle tracking rates that may enable a new generation of simulation-based design and engineering work that was previously impractical to perform.

\section*{Acknowledgements}
This research was supported by the Exascale Computing Project (17-SC-20-SC), a collaborative effort of the U.S. Department of Energy Office of Science and the National Nuclear Security Administration. The material was based upon work supported by the U.S. Department of Energy, Office of Science, under contract DE-AC02-06CH11357. This research used resources of the Argonne Leadership Computing Facility, which is a U.S. Department of Energy Office of Science User Facility operated under contract DE-AC02-06CH11357. This research used resources of the Oak Ridge Leadership Computing Facility at the Oak Ridge National Laboratory, which is supported by the Office of Science of the U.S. Department of Energy under Contract No. DE-AC05-00OR22725. We gratefully acknowledge the computing resources provided and operated by the Joint Laboratory for System Evaluation (JLSE) at Argonne National Laboratory. Intel performance studies were done on a pre-production supercomputer with early versions of the Aurora software development kit.

\bibliography{bibliography}

\begin{thebibliography}{20}

\bibitem{Tramm2022}
J.R. Tramm, P.K. Romano, J.~Doerfert, A.L. Lund, P.C. Shriwise, A.R. Siegel, G.~Ridley, A.~Pastrello, \emph{Toward Portable {GPU} Acceleration of the {OpenMC Monte Carlo} Particle Transport Code}, in \emph{PHYSOR 2022 - International Conference on Physics of Reactors} (2022)

\bibitem{hamilton19annals}
S.~Hamilton, T.~Evans, \emph{Continuous-energy {M}onte {C}arlo neutron transport on {GPU}s in the {S}hift code}, Ann. Nucl. Energy \textbf{128}, 236 (2019)

\bibitem{choi}
N.~Choi, K.M. Kim, H.G. Joo, \emph{Optimization of neutron tracking algorithms for gpu-based continuous energy monte carlo calculation}, Annals of Nuclear Energy \textbf{162} (2021)

\bibitem{Mckinley2019}
M.S. Mckinley, R.~Bleile, P.S. Brantley, S.~Dawson, M.O. Brien, M.~Pozulp, D.~Richards, \emph{{Status of LLNL Monte Carlo Transport Codes on Sierra GPUs}}, in \emph{International Conference on Mathematics and Computational Methods Applied to Nuclear Science and Engineering (M{\&}C 2019)} (2019), pp. 2160--2165

\bibitem{romano-2015-ane1}
P.K. Romano, N.E. Horelik, B.R. Herman, A.G. Nelson, B.~Forget, \emph{{OpenMC}: A state-of-the-art {Monte} {Carlo} code for research and development}, Ann. Nucl. Energy \textbf{82}, 90 (2015), \url{https://doi.org/10.1016/j.anucene.2014.07.048}

\bibitem{Romano2012}
P.K. Romano, B.~Forget, \emph{Parallel fission bank algorithms in monte carlo criticality calculations}, Nuclear Science and Engineering \textbf{170}, 125 (2012)

\bibitem{ridley2021}
G.~Ridley, B.~Forget, \emph{Design and {{Optimization}} of {{GPU Capabilities}} in {{OpenMC}}}, Trans. Am. Nucl. Soc. \textbf{125}, 456 (2021)

\bibitem{openmc_optix}
J.~Salmon, S.~McIntosh-Smith, \emph{Exploiting Hardware-Accelerated Ray Tracing for Monte Carlo Particle Transport with OpenMC}, in \emph{2019 IEEE/ACM Performance Modeling, Benchmarking and Simulation of High Performance Computer Systems (PMBS)} (2019), pp. 19--29, \urlstyle{tt}\url{https://doi.org/10.1109/PMBS49563.2019.00008}

\bibitem{openmc_exasmr}
P.K. Romano et~al., \emph{The {ExaSMR OpenMC Monte C}arlo particle transport code repository} (2023), \urlstyle{tt}\url{https://github.com/exasmr/openmc}

\bibitem{openmc_main}
P.K. Romano et~al., \emph{{OpenMC Monte Carlo} particle transport code repository} (2023), \urlstyle{tt}\url{https://github.com/openmc-dev/openmc}

\bibitem{serpent}
J.~Leppänen, M.~Pusa, T.~Viitanen, V.~Valtavirta, T.~Kaltiaisenaho, \emph{The serpent monte carlo code: Status, development and applications in 2013}, Annals of Nuclear Energy \textbf{82}, 142 (2015)

\bibitem{mcnp_62}
C.J. Werner, J.S. Bull, C.J. Solomon, F.B. Brown, G.W. McKinney, M.E. Rising, D.A. Dixon, R.L. Martz, H.G. Hughes, L.J. Cox et~al., Tech. rep. (2018), \urlstyle{tt}\url{https://www.osti.gov/biblio/1419730}

\bibitem{Kowalski2021}
M.A. Kowalski, P.~Cosgrove, J.~Broman, E.~Shwageraus, \emph{{S}{C}{O}{N}{E}: {A} {S}tudent-{O}riented {M}odifiable {M}onte {C}arlo {P}article {T}ransport {F}ramework}, Journal of Nuclear Engineering \textbf{2}, 57 (2021)

\bibitem{Leppanen2009}
J.~Lepp{\"{a}}nen, \emph{{Two practical methods for unionized energy grid construction in continuous-energy Monte Carlo neutron transport calculation}}, Annals of Nuclear Energy \textbf{36}, 878 (2009)

\bibitem{ecp-se-08-43}
K.~Smith, Milestone Report WBS 1.2.1.08 ECP-SE-08-43, Exascale Computing Project (2017)

\bibitem{forget2014}
B.~Forget, S.~Xu, K.~Smith, \emph{Direct {D}oppler broadening in {Monte Carlo} simulations using the multipole representation}, Annals of Nuclear Energy \textbf{64}, 78 (2014)

\bibitem{forget2022}
B.~Forget, J.~Yu, G.~Ridley, \emph{Performance Improvements of the Windowed Multipole Formalism Using a Rational Fraction Approximation of the {F}addeeva Function}, in \emph{PHYSOR 2022 - International Conference on Physics of Reactors} (2022)

\bibitem{romano2018}
P.K. Romano, J.A. Walsh, \emph{An improved target velocity sampling algorithm for free gas elastic scattering}, Annals of Nuclear Energy \textbf{114}, 318 (2018)

\bibitem{ECP-SE-08-119}
S.~Hamilton, T.~Evans, P.~Romano, J.~Tramm, E.~Merzari, M.~Min, P.~Fischer, Milestone Report WBS 2.2.2.03 ECP-SE-08-119, Exacale Computing Project (2022)

\bibitem{techpowerup}
\emph{Techpowerup} (2023), accessed on April 4, 2023, \urlstyle{tt}\url{https://www.techpowerup.com}

\end{thebibliography}

\pagestyle{empty}
\vspace*{\fill}
\noindent\fbox{%
  \parbox{\textwidth}{%
    The submitted manuscript has been created by UChicago Argonne, LLC, Operator
    of Argonne \mbox{National} Laboratory (``Argonne''). Argonne, a U.S.
    Department of Energy Office of Science laboratory, is operated under
    Contract No. \mbox{DE-AC02-06CH11357}. The U.S. Government retains for
    itself, and others acting on its behalf, a paid-up nonexclusive, irrevocable
    worldwide license in said article to reproduce, prepare derivative works,
    distribute copies to the public, and perform publicly and display publicly,
    by or on behalf of the Government. The Department of Energy will provide
    public access to these results of federally sponsored research in accordance
    with the DOE Public Access Plan.\\
    \url{https://energy.gov/downloads/doe-public-access-plan}}%
}
\vspace*{\fill}

\end{document}